\documentclass[aps,prl,letterpaper,11pt,twoside,tightenlines,nofootinbib,showpacs,preprint,twocolumn]{revtex4}
\usepackage{graphicx}
\usepackage[sort&compress]{natbib}
\usepackage{latexsym}
\usepackage{epsfig}
\usepackage{amsmath}
\begin{document}
\newcommand{\eg}{{\it e.g.}}
\newcommand{\etal}{{\it et. al.}}
\newcommand{\ie}{{\it i.e.}}
\newcommand{\be}{\begin{equation}}
\newcommand{\dd}{\displaystyle}
\newcommand{\ee}{\end{equation}}
\newcommand{\bea}{\begin{eqnarray}}
\newcommand{\eea}{\end{eqnarray}}
\newcommand{\bef}{\begin{figure}}
\newcommand{\eef}{\end{figure}}
\newcommand{\bce}{\begin{center}}
\newcommand{\ece}{\end{center}}
\def\lsim{\mathrel{\rlap{\lower4pt\hbox{\hskip1pt$\sim$}}
    \raise1pt\hbox{$<$}}}         
\def\gsim{\mathrel{\rlap{\lower4pt\hbox{\hskip1pt$\sim$}}
    \raise1pt\hbox{$>$}}}         

\title{QCD Equation of State and Cosmological Parameters in Early Universe}
\author{P.~Castorina$^{a,b}$, V. Greco$^{a,c}$ and S. Plumari$^{a,c}$}

\affiliation{$^a$Dipartimento di Fisica, Universit\`a di Catania,Via Santa Sofia 64
I-95123 Catania, Italia}
\affiliation{$^b$INFN Sezione di Catania, Via Santa Sofia 64
I-95123 Catania, Italia}
\affiliation{$^c$INFN Laboratori Nazionali del Sud, Via Santa Sofia 62, I-95123 Catania, Italy}

\date{\today}
\begin{abstract}
The time evolution of cosmological parameters in early Universe at the deconfinement transition is studied by an equation of state (EoS) which takes into account the finite baryon density and the background magnetic field. The non perturbative dynamics is described by the Field Correlator Method (FCM)  which gives, with a small number of free parameters, a good fit of lattice data. The entire system has two components, i.e. the quark-gluon plasma and the electroweak sector, and the solutions of the Friedmann equation show that the scale factor, $a(t)$, and $H(t)= (1/a)da/dt$ are weakly dependent on the EoS, but the deceleration parameter, $q(t)$, and the jerk, $j(t)$, are strongly modified above the critical temperature $T_c$, corresponding to a critical time $t_c \simeq 20-25 \mu s$. The time evolution of the cosmological parameters suggest that above and around $T_c$ there is a transient state of acceleration typical of a matter dominated
Universe; this is entailed by the QCD strong interaction driven by the presence of massive colored objects.

 
\end{abstract}
 \maketitle

\section{Introduction}

Lattice simulations of Quantum Chromodynamics (QCD) and  phenomenological analyses of relativistic heavy ion collisions data clearly indicate that the transition from quarks and gluons to colorless states occurs in a non perturbative regime such that
for large baryon chemical potential, $\mu_B$, the transition is first order and for small $\mu_B$ it is cross-over  at a  (pseudo) critical temperature $T_c \simeq 155 \pm 15$ Mev \cite{fodor,karsch}. 

Moreover  lattice QCD in a background magnetic field , $B$, shows \cite{katz,Bonati:2013vba} that
the transition temperature is reduced by the magnetic field and that  the
thermal QCD medium is paramagnetic around and above the transition temperature.

The previous parameters, $T,\mu_B$ and $B$, give different and  important informations on the dynamics of the phase transition which, in turn,  has strong implications at cosmological level in the early Universe ( see \cite{boe1,bom} for a recent review).

In particular, the equation of state (EoS) in the quark-gluon plasma phase affects directly the density fluctuactions of the thermodynamic quantities 
\cite{gw4,flor,greco,brazil} and  the emission of gravitational waves \cite{gw4,gw1,gw2,gw3,gw5,gw6} during the cosmological evolution.

In ref. \cite{gw4} the density fluctuactions in the early Universe have been studied by using the MIT bag model EoS and the old lattice QCD EoS with a first order deconfinement phase transition.
A more recent and realistic QCD EoS has been introduced in ref. \cite{flor} improving the reliability of the calculations of the fluctuactions and of the modifications in the time evolution of the cosmological scale factor $a(t)$.  
The effects of the external magnetic field on the energy density fluctuations have been considered in ref.\cite{greco} by a phenomenological QCD EoS which includes $B$. An analogous analysis for finite density $\mu_B$  with different variants of MIT bag model has been carried out in ref. \cite{brazil}.

The results of the previous studies show a smooth time dependence of the thermodynamic quantities, however they essentially  consider the effect of the transition on the thermodynamic fluctuactions and on the scale factor. 

In this paper we shall analize the modifications on the early Universe evolution, by using a realistic EoS which depends on $T,\mu_B$ and $B$, with a particular focus on the evolution of the cosmological parameters rather than on the fluctuactions.

To describe the EoS in the quark-gluon plasma phase we shall apply the Field Correlator Method (FCM) \cite{simo}. This specific choice is motivated by the non perturbative dynamics of the FCM which describes, with a small number of free parameters, lattice data of the thermodynamics quantities ( pressure and energy density) at finite temperature and their dependence on $\mu_B$ and $B$ \cite{fcm2,fcm3,fcm4}.
Here we will also show that its extension to a finite B \cite{fcmB1,fcmB2} capture correctly the
main features of lQCD EoS under a magnetic field.

The discussion of the combined effects of the chemical potential and of the external magnetic field, during the deconfinement transition, in the Friedmann equations is of interest in its own right. However, as we shall see, the most interesting results are on the behavior of the deceleration parameters, $q$, and the jerk, $j$, defined as ($'$ indicates the time derivative)
\be
q= - \frac{a^{''} a}{a^{'2}}
\ee
\be
j= \frac{a^{'''} a^2}{a^{'3}},
\ee

which have an important role in describing the cosmological evolution \cite{gibbons,vari}.

Indeed, it turns out that the deceleration and the jerk  are strongly modified above the critical temperature $T_c$, corresponding to a critical time $t_c \simeq 20-25 \mu s$, and that the EoS and the time evolution of the cosmological parameters suggest that above $T_c$ the system has the typical behavior of a matter dominated Universe, with clusters of colored particles.  

The plane of the paper is as follows. In Sec. 1 we shall consider the relevant cosmological equations and parameters. The FCM, with the dependence on $\mu_B$ and $B$ is recalled and compared with lattice data in Sec.2. Sec.3 is devoted to the solution of the Friedmann equation with the FCM EoS and Sec.4 contains the electroweak contributions to the EoS of the entire dynamical system. In Sec.5 the results on the cosmological parameters are discussed and  our comments and conclusions are in Sec.6.

\section{1. Cosmological parameters}

The parameters $H(t)=a^{'}/a$, $q(t)$and $j(t)$ can naturally be defined making use of
Taylor series for the scale factor $a(t)$ near a generic time $t^*$:
\begin{multline}
a(t)=a(t^*)+a^{'}(t^*)(t-t^*)+\frac{1}{2}a^{''}(t^*)(t-t^*)^2\\
+\frac{1}{3!}a^{'''}(t^*)(t-t^*)^3+...
\end{multline}
which can be written as
\begin{multline}
a(t) = a(t^*)[1 +H(t^*)(t-t^*)-\frac{1}{2} (qH^2)(t^*)(t-t^*)^2 \\
+\frac{1}{3!}(jH^3)(t^*)(t-t^*)^3+...
\end{multline}
Basic characteristics of the cosmological evolution, both static and
dynamical, can be expressed in terms of $H_0$, the present time value of $H(t)$, and of the deceleration $q_0$. The other
parameters, i.e. the higher time derivatives of the scale factor, enable to construct model-independent kinematics of the cosmological expansion.

Indeed cosmological models can be tested by expressing the Friedmann equation in terms of directly measurable
cosmological scalars constructed out of higher derivatives of the scale factor, i.e $q, j,..$ \cite{gibbons,vari}.

To illustrate this aspect, let us consider a simple two-component Universe filled with non-relativistic matter with density $M_m/a^3$ and radiation with density $M_r/a^4$ ( $M_m, M_r$ constants)  which do not interact with each other \cite{vari} and without a magnetic background. By writing the Friedmann
equation in the form ($8\pi G/3 =1$)
\be
\frac{a^{'2}}{a^2}+\frac{k}{a^2} = \frac{M_m}{a^3} + \frac{M_r}{a^4}
\ee
and diffrentiating twice with respect to $t$, one gets
\be
a^{''} = - \frac{1}{2} \frac{M_m}{a^2} - \frac{M_r}{a^3}
\ee
and
\be
a^{'''} =  \frac{M_m}{a^3} a^{'} + 3 \frac{M_r}{a^4} a^{'}. 
\ee
Then the deceleration and the jerk can be written as $q=A/2+R$ and $j=A+3R$ where $A=M_m/a^3H^2$ and $R=M_r/a^4H^2$. For a flat Universe filled only with non relativistic matter one has $R=0$, $q=1/2$, $A=1$ and $j=1$; if one considers only radiation,then $A=0$, $q=1$, $R=1$ and $j=3$. A deviation from these values of $q$ and $j$ indicates a mixture of matter and radiation and/or an interaction between the two components.Indeed the same values can be easily obtained by a simple application of the Friedmann equation 
(for $k=0$), written as
\be
\frac{a'}a{}=-\frac{d\epsilon/dt}{3(\epsilon+p)}= \sqrt{\frac{8 \pi G}{3}} \sqrt{\epsilon},
\ee
by the EoS $p=c_s^2 \,\epsilon$ with a constant speed of sound. The analitic solution is
\be
\epsilon(t)=\frac{4 \, \epsilon(t_0)}{\big[ 3\sqrt{\frac{8 \pi G \, \epsilon(t_0)}{3}}(1+c_s^2)(t-t_0)+2\big]^2}
\ee
and
\be
\frac{a(t)}{a(t_0)}=\big[\frac{3}{2}\sqrt{\frac{8 \pi G \, \epsilon(t_0)}{3}}(1+c_s^2)(t-t_0)+1\big]^{\frac{2}{3(1+c_s^2)}}
\ee
and one obtains
\be
q=-\frac{a'' a}{a'^2}=\frac{1+3 c_s^2}{2}
\ee
\be
 j=\frac{a''' a^2}{a'^3}=\big(\frac{1+3 c_s^2}{2}\big)(2+3c_s^2)
\ee
which reproduce the previous values  for matter dominated ($c_s^2=0$) and radiation dominated ($c_s^2=1/3$) Universe.

In general the speed of sound is not constant and in the next sections we shall discuss the behavior of the cosmological parameters $q$ and $j$ during the deconfinement transition on the basis of the Friedmann equation (8)
and by using  the energy density, $\epsilon$ and the pressure, $p$, in the FCM, after fitting the QCD lattice data at finite temperature, $\mu_B$ and $B$.

\section{2.Field Correlator Method}

Many phenomenological models of QCD at finite temperature and density  cannot make reliable 
predictions for the two relevant limits, i.e. high temperature and small
chemical potential or high chemical potential and low
temperature. This is clearly a serious drawback, since
those models cannot be fully tested. One of the few exceptions is the Field Correlator Method (FCM)\cite{simo}, which  is able to cover the full temperature-chemical
potential-magnetic background field space and contains ab initio
the property of confinement, which is expected to
play a role, at variance with other models like, e.g., the
Nambu - Jona Lasinio model.

Indeed, the
approach based on the FCM provides a systematic tool to treat non perturbative effects
in QCD  by gauge invariant field correlators and gives a natural treatment
of the dynamics of confinement (and of the deconfinement
transition) in terms of the Gaussian, i.e. quadratic in the tensor $F_{\mu \nu}$, correlators for the chromo-electric (CE) field, $D^E$ and $D^E_1$, and  for the chromo-magnetic field (CM),  $D^H$ and $D^H_1$.
 
In particular, these correlators are related to the simplest
non trivial 2-point correlators for the CE and CM fields
by
\begin{multline}
g^2 < Tr_f[C_i(x) \Phi(x,y) C_k(y) \Phi(y,x)] > =\\
\delta_{ik}[D^C(z) + D^C_1(z) +z_4^2 \frac{\partial D^C_1(z)}{\partial z^2}]\\
\pm z_i z_k \frac{\partial D^C_1(z)}{\partial z^2}
\end{multline}
where $z=x-y$ and $C$  indicates the CE (E) field or CM
(H) field (the minus sign in the previous expression corresponds to the magnetic case) and
\be
\Phi= P exp[ig \int_x^y A^\mu dz_\mu]
\ee
is the parallel transporter.

The FCM has been extended to finite temperature
 and chemical potential  \cite{fcm2,fcm3,fcm4} and 
the analytical results, in the gaussian approximation, are in good agreement with  the lattice data on thermodynamic quantities,
 (available for small $\mu_B$ only). Moreover, the application  of the FCM for
large values of the chemical potential allows to obtain a
simple expression of the Equation of State of the quark-
gluon matter in the  range of the baryon density relevant for the study of neutron stars \cite{we1,we2,we3}.

The comparisons with lattice data of the FCM predictions for the pressure ($p/T^4$),the interaction measure, $\Delta/T^4=(\epsilon-3p)/T^4$,  and the speed of sound, at $\mu_B =0$ and $\mu_B =0.4 \rm GeV$,  are depicted in figs. 1,2 and 3.

\begin{figure}
{{\epsfig{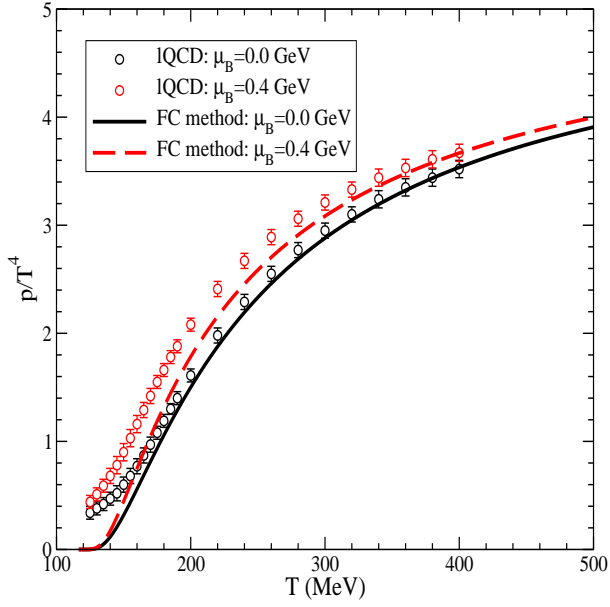}}
\caption{Comparison with lattice data \cite{fodor,karsch,lattice1} of the pressure, $p$, evaluated by the FCM.  
}}
\end{figure}

\begin{figure}
{{\epsfig{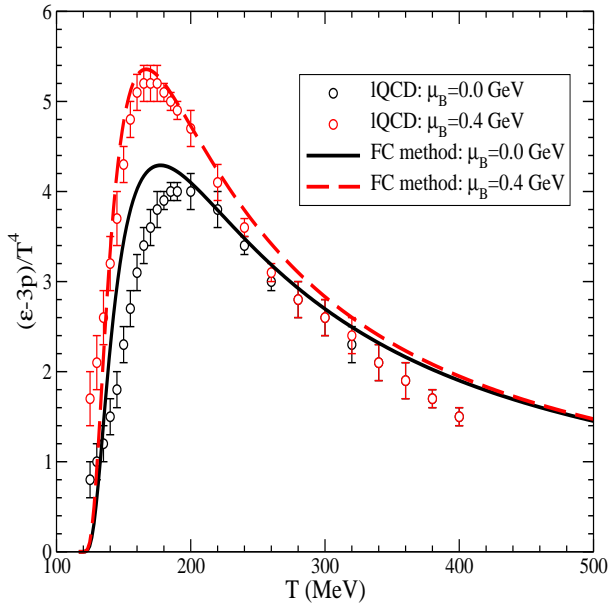}}
\caption{Comparison with lattice data \cite{fodor,karsch,lattice1} of the interaction measure $\Delta=(\epsilon-3p)/T^4$ evaluated by the FCM.  
}}
\end{figure}

\begin{figure}
{{\epsfig{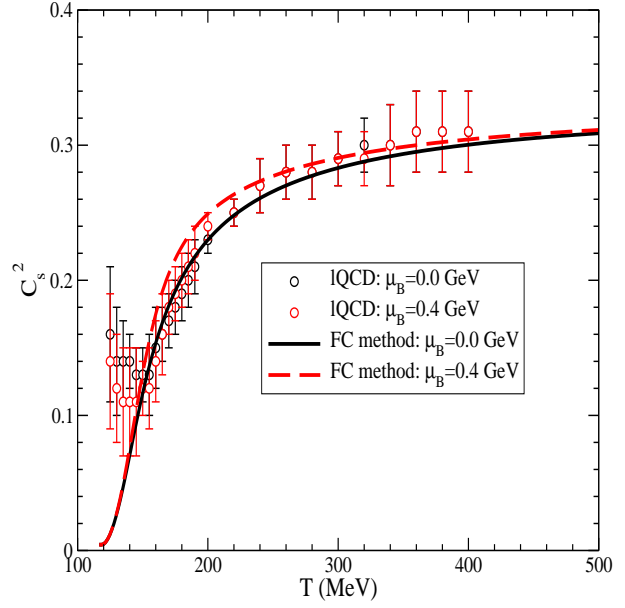}}
\caption{Comparison with lattice data \cite{fodor,karsch,lattice1} of the speed of sound evaluated by the FCM.  
}}
\end{figure}

More recently,  the effect of a background magnetic field, $B$, has been included in the FCM equation of state \cite{fcmB1,fcmB2} and the quark (q) pressure and the gluon (g) pressure  turn out to be

\bea
p_q (B) &=& \frac{N_ce_q B}{2\pi^2} \big[ \phi(\mu) + \phi(-\mu) \nonumber \\ 
& & +\frac{2}{3} \frac{(\lambda (\mu) +\lambda (-\mu))}{e_qB} \nonumber \\ 
& &-\frac{e_q B}{24} (\tau(\mu)+ \tau(-\mu))  \big], \label{PeB0}
\eea
\be
p_g=\frac{N_c^2-1}{3 \pi^2} T^4 \int^\infty_0 \frac{z^3 dz}{\exp{\left( z+\frac{9}{4}\frac{V_1(T)}{2T} \right)}-1}
\ee 
where $\phi(\mu)$ ,$\lambda(\mu)$ and $\tau(\mu)$ are respectively given by 
\be
\phi(\mu) = \int^\infty_0 \frac{p_zdp_z}{1+\exp{\left( \frac{p_z-\bar \mu}{T}\right)}} 
\ee
\be
\lambda(\mu) = \int^\infty_0 \frac{p^4dp}{\sqrt{p^2+ \tilde m^2_q}}
\frac{1}{\exp \left( \frac{\sqrt{ p^2+\tilde m^2_q}-\bar
\mu}{T}\right)+1},\label{P_eB}
\ee
\be
\tau(\mu) = \int^\infty_0 \frac{dp_z}{\sqrt{p_z^2+ \tilde m^2_q}} \frac{1}{\exp
\left( \frac{\sqrt{ p^2_z+\tilde m^2_q}-\bar \mu}{T}\right)+1}, 
\ee
with $\bar \mu=\mu - \frac{V_1(T)}{2}$, $\tilde m^2_q= m^2_q+e_qB$ with $m_q$ the current mass
($m_d= 5 \rm GeV$, $m_u=10 \rm MeV$ and $m_s= 140 \rm MeV$), and
\bea
V_1(T)&=&c_0 + f(T/T_c)\nonumber\\
&=&0.15 + 0.175 \,\left(  1.35 \frac{T}{T_C} -1\right) ^{-1}
\eea
is the quark-quark interaction potential.

In the limit $eB\to 0$ the only non zero terms are $\lambda(\mu)$ and $\lambda(-\mu)$ and the previous formulas 
reproduce the case $\mu_B \ne 0$ \cite{fcm2}.

 In fig. 4 one compares the ratio $s/T^3$ ($s$ being the entropy density) evaluated in the FCM with lattice data for different values of the background field 
 \cite{lattice1}. 

\begin{figure}
{{\epsfig{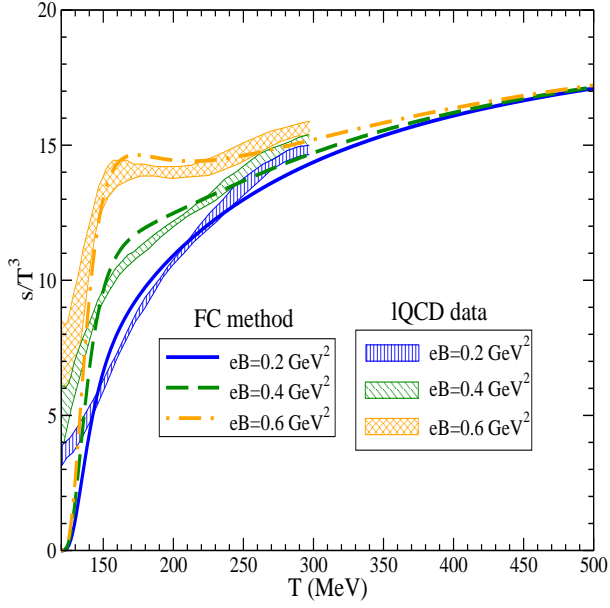}}
\caption{Comparison of the FCM calculations for the entropy density with lattice data \cite{katz} for QCD in a background magnetic field $B$  
}}
\end{figure}

We consider a value of $eB=0.6 \rm GeV^2$ that corresponds to a maximum estimate of the magnetic field
generated at the early times \cite{vacha}. In some possible scenario also a finite $\mu_B$
could be still significant ($\mu_B\approx T$) at the QCD transition \cite{bom}. We consider a $\mu_B=0.4 \,\rm GeV$,  but we will see that its impact is however quite limited.

The deconfinement temperature depends on $\mu_B$ and $B$ and therefore the critical time $t_c$ of the transition
turns out to be $24 \,\mu s$, $21.5 \,\mu s$, $15.6 \,\mu s$ and $13.6 \,\mu s$ respectively for $\mu_B=0.0 \rm GeV$ and $B=0.0$, $\mu_B=0.4 \rm GeV$ and $B=0.0$, $\mu_B=0.0$ and $B=0.6 \rm GeV^2$, and $\mu_B=0.4 \rm\, GeV$ and $B=0.6 \, \rm GeV^2$.

\section{3.Friedmann equation and quark-gluon plasma in the FCM}

In the FCM the EoS of the quark-gluon plasma, i.e. $p(\epsilon)$, is obtained by direct calculations of $p(T)$ and $\epsilon(T)$, which inserted in  the Friedman equation (8) give  the time dependence of the temperature , $T(t)$, and the corresponding time evolution of the thermodynamic quantities.

For the initial conditions $t_i = 1\mu s$, $T_i \simeq 500$ MeV and $\epsilon_i \simeq 110 $ GeV/$fm^3$, the function $T(t)$, solution of the Friedman equation, is shown in fig.5 for different values of $\mu_B$ and compared with the MIT bag model with the same initial conditions and a bag pressure $B_{mit}= 220$ MeV/fm$^3$. 

The arrows in fig.5 (and in the next figures) correspond to the critical temperatures for the different  specific sets  of the parameters and therefore to the corresponding critical time, $t_c$, when the phase transition occurs. The critical  temperature $T_c = 160$ MeV, for $\mu_B=0$,corresponds to $t_c \simeq 25 \mu s$, which decreases to $t_c \simeq 22 \mu s$ for $\mu_B=0.4$ GeV. The curves are also plotted for $t > t_c$, i.e. for temperature below the transition point although the effective degrees of freedom below $T_c$ are non included in the present paper. 

The equation of state $p/\epsilon$ in the FCM, reported in fig. 6, shows a small dependence on $\mu_B$ up to $400$ MeV.

\begin{figure}
{{\epsfig{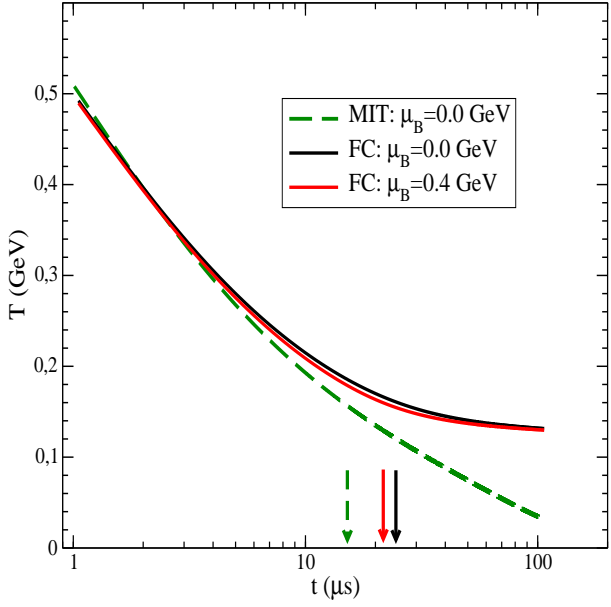}}
\caption{$T(t)$ solution of the Friedmannn equation (8) by using the EoS in the FCM. The arrows indicate the critical time, $t_c$ of the transition given in the text.   
}}
\end{figure}

\begin{figure}
{{\epsfig{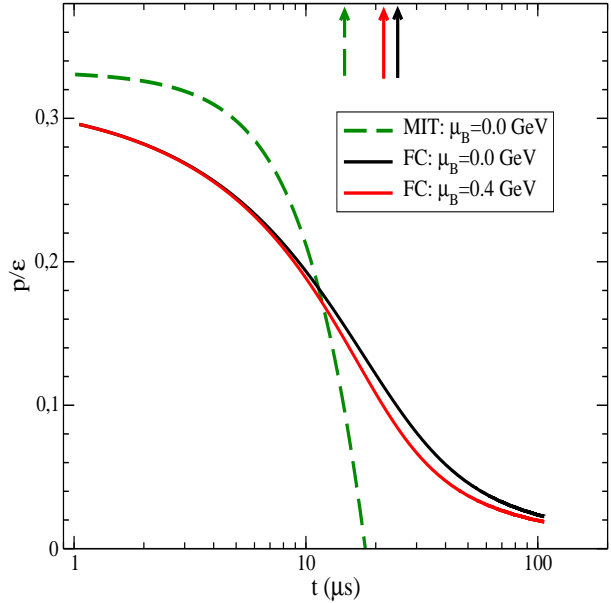}}
\caption{EoS in the FCM for finite chemical potential. The green curve referes to the MIT bag model calculation with the parameters given in the text.   
}}
\end{figure}

The role of the background magnetic field in the calculation of $T(t)$ and in the time evolution of the EoS are reported in fig.7 and in fig.8, which also describe the combined effect of the finite density and of the background magnetic field. 

\begin{figure}
{{\epsfig{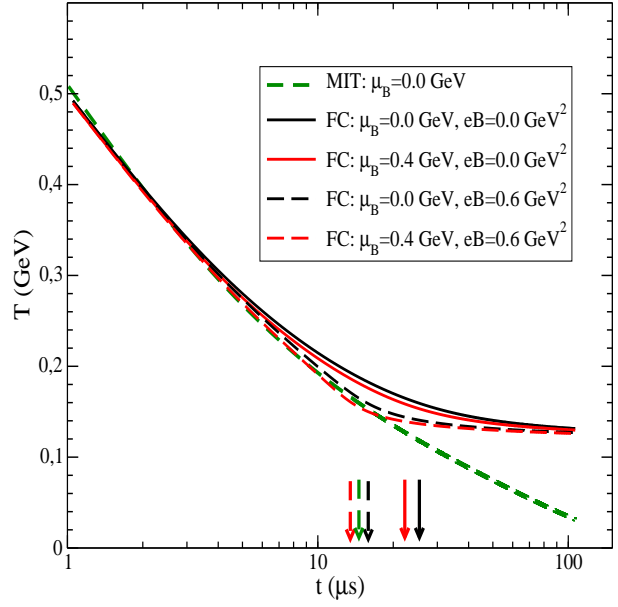}}
\caption{The temperature profile $T(t)$ for different values of $\mu_B$  and $B$.  
}}
\end{figure}

\begin{figure}
{{\epsfig{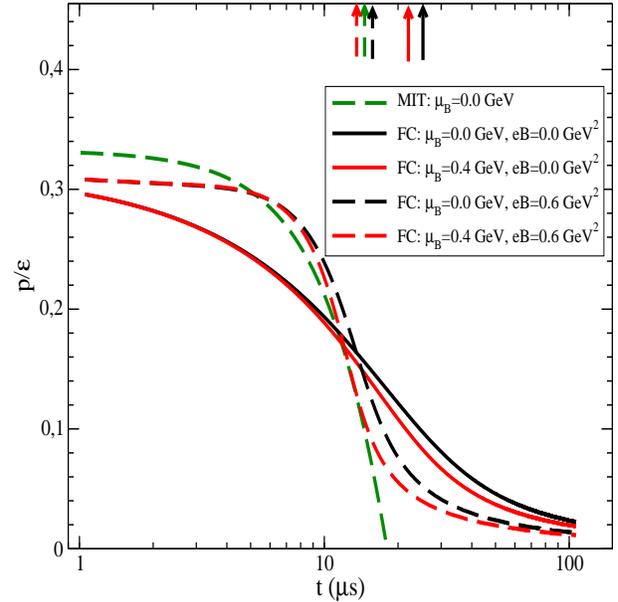}}
\caption{Time evolution of the EoS in the FCM for finite chemical potential and background magnetic field. 
}}
\end{figure}

Notice that the results in the MIT bag model ( green curves in figs. 6,7,8) are clearly different from the EoS evaluated by the FCM and that the magnetic field B can move down $t_c$ by nearly a factor of 2.

\section{4.Electroweak contribution to the Equation of State}

The time evolution of the cosmological parameters depend on the EoS of the entire systems and therefore one has to take into account not only the quark-gluon plasma (qgp) degrees of freedom, discussed in the previous section,  but also  the contributions of the electroweak sector to the pressure, $p_{ew}$ and the energy density, $\epsilon_{ew}$. The total pressure and energy density are therefore
 $p_{tot}=p_{qgp}+p_{ew}$ and $\epsilon_{tot}=\epsilon_{qgp}+\epsilon_{ew}$.

The electroweak sector is described as a system of free massless particles, i.e.
\be
\epsilon_{ew}=g_{ew}\frac{\pi^2}{30}T^4,
\ee
\be
p_{ew}=g_{ew}\frac{\pi^2}{90}T^4,
\ee
with the number of degrees of freedom $g_{ew}=14.45$.

The temperature profile is almost unmodified by the introduction of the electroweak contributions as one can see from fig.9, where the curves are plotted for $\mu_B=0$, since the chemical potential produces negligible changes (see figs. 5,6).

\begin{figure}
{{\epsfig{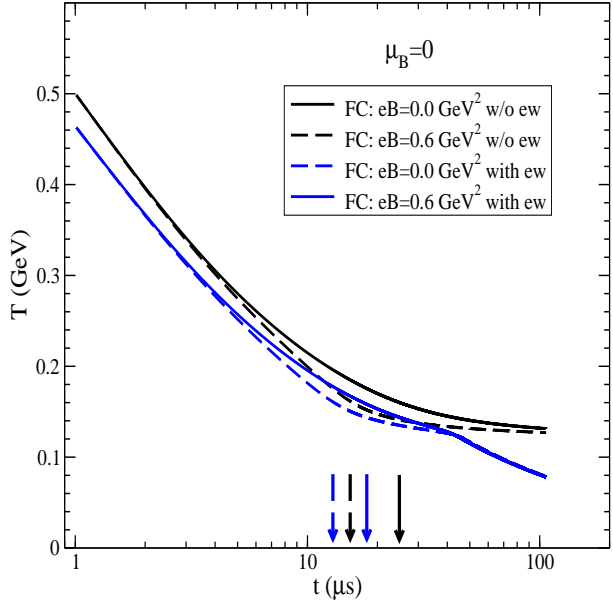}}
\caption{$T(t)$ with/without the electroweak contribution and the background magnetic field. 
}}
\end{figure}

On the other hand, the EoS of the entire system strongly reflects the presence of the electroweak terms.
In fig. 10 the time evolution of $p/\epsilon$ for the entire system is depicted and, indeed,  one immediately notes that in the time interval $t \simeq 10-25 \mu s$ the electroweak part has a minor role, but near the critical temperature there is a clear change in the shape and the EoS tends to $p=\epsilon/3$, i.e. to a  radiation dominated Universe, for long time. The curves are plotted for time longer than  $t_c$,  just to show the behavior below $T_c$ without the contribution of the colorless effective degrees of freedom after the transition, not included in the present analysis and that will be discussed in a forthcoming paper.  

\begin{figure}
{{\epsfig{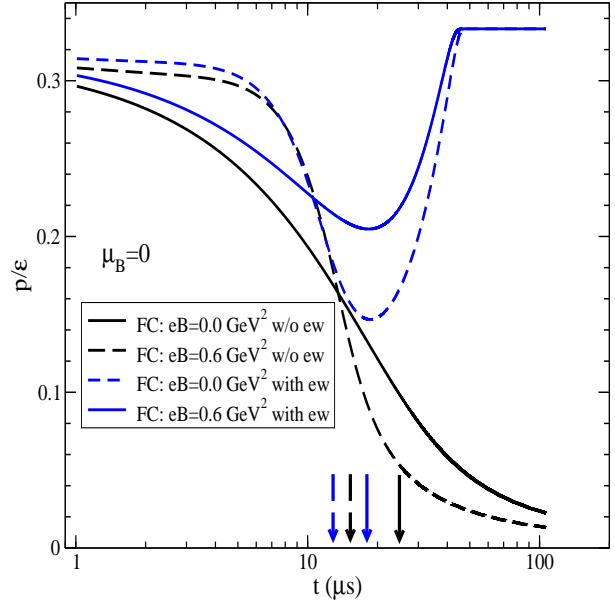}}
\caption{EoS with/without the electroweak contribution and the background magnetic field, for $\mu_B=0$ 
}}
\end{figure}

\section{5.Evolution of the cosmological parameters} 

The evolution of the cosmological parameters during the deconfinement transition is directly related with the EoS.
In fig. 11 and in fig. 12 are respectively depicted the time behavior of the scale factor and of $H(t)$ for different values of the chemical potential, of the magnetic field and also with/without the electroweak contribution. The final result is essentially independent on the specific setting.

\begin{figure}
{{\epsfig{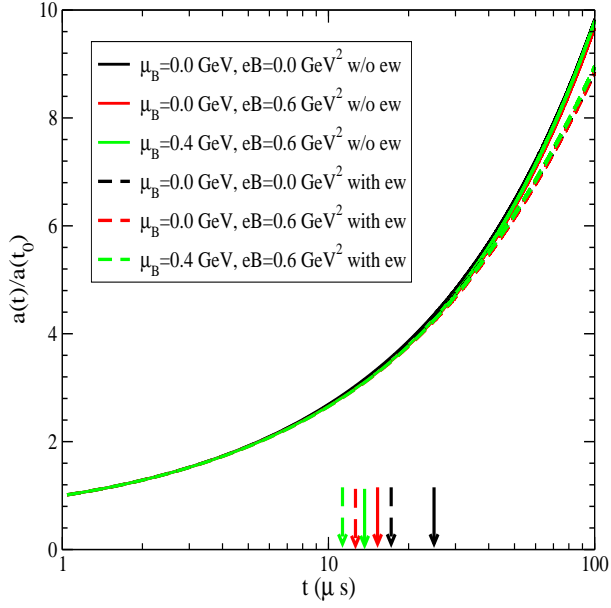}}
\caption{Time evolution of the scale factor for finite chemical potential, with/without the electroweak contribution and the background magnetic field.
}}
\end{figure}

\begin{figure}
{{\epsfig{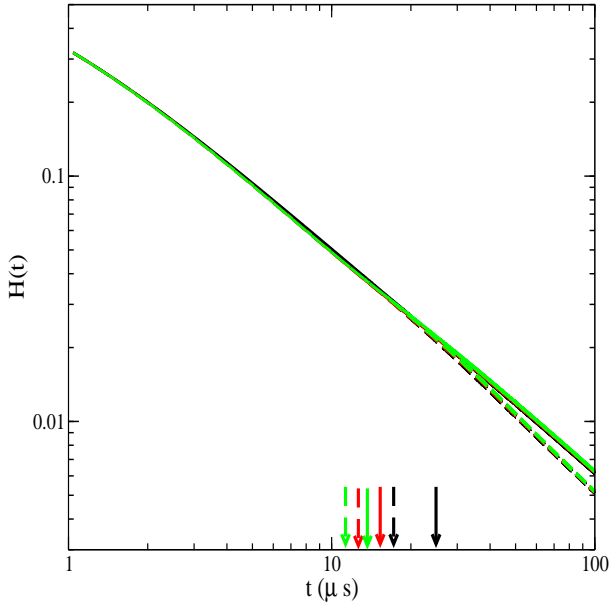}}
\caption{$H(t)$ for finite chemical potential, with/without the electroweak contribution and the background magnetic field. 
}}
\end{figure}

However the deceleration $q(t)$ and the jerk $j(t)$ strongly follows the time evolution of the EoS (in fig. 10). Both $q$ and $j$ ( see fig. 13 and fig. 14) after  initial values corresponding to a radiation dominated Universe ($q \simeq 1$, $j \simeq 3$), tend to  $q \simeq 1/2$ and $j \simeq 1$, typical of a matter dominated Universe, approaching from above the critical temperature. Indeed, without the electroweak sector  the values of the cosmological parameters for time $t \simeq 100 \mu s$ would be the typical ones of a matter dominated Universe. 
However , near the transition point the electroweak terms in the EoS start to be relevant  and therefore $q \rightarrow 1$ and $j \rightarrow 3$ for long time.

\begin{figure}
{{\epsfig{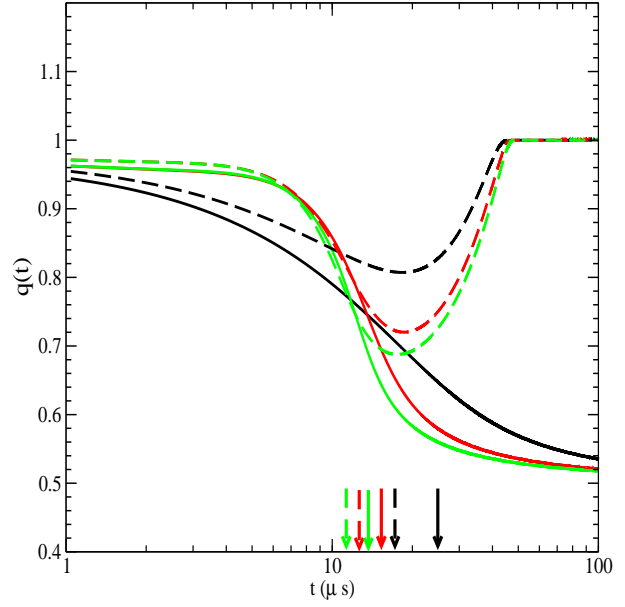}}
\caption{$q(t)$ for finite chemical potential, with/without the electroweak contribution and the background magnetic field. 
}}
\end{figure}

\begin{figure}
{{\epsfig{file=jerk.eps,height=8.0 true cm,width=8.0 true cm, angle=0}}
\caption{$j(t)$ for finite chemical potential, with/without the electroweak contribution and the background magnetic field. 
}}
\end{figure}

\begin{figure}
{{\epsfig{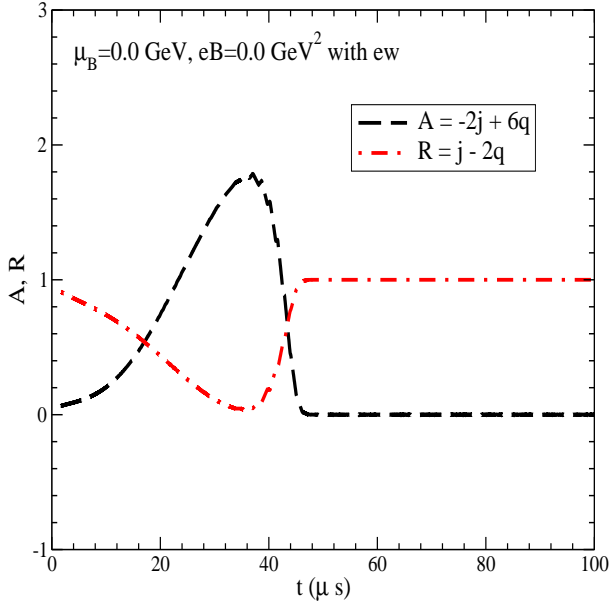}}
\caption{Time evolution of $A$ and $R$ according to the simple model in eq.(5) but using $q(t)$ and $j(t)$ in the FCM. 
}}
\end{figure}

A possible interpretation of this peculiar behavior can be done with the help of the simple model in Sec.1, where 
$A= -2j+6q$ and $R=j-2q$ are respectively the matter and radiation terms in the Friedmann equation.  For a matter dominated flat Universe $A=1$ and $R=0$ whereas for a radiation dominated one  $A=0$ and $R=1$ and since in eq.(5) there is no  interaction between the two components $A+R=1$. Of course, this condition in not satisfied by using  $q$ and 
$j$ calculated in the FCM and different values of $A$ and $R$ signal a mixture and/or an interaction between the two components. 

By using the values of $q$ and $j$ in the FCM, the time evolution of $A$ and $R$ is given in fig. 15. $R$ decreases from an initial value $\simeq 1$ to $\simeq 0.3$ at $t \simeq 25 \mu s$ and to $\simeq 0$ at $t \simeq 32\mu s$ and then quickly reaches again the value of a radiation dominated Universe. $A$ has the corresponding evolution , starting from $A\simeq 0.0$, increasing to a value $\simeq 1$ at $t \simeq 25 \mu s$, with a maximum  $A \simeq 1.8$  at $t \simeq 35 \mu s$, and finally decreasing to $A \simeq 0$,i.e. a radiation dominated Universe.

Therefore there is a clear mixture of the two components which strongly interact each other.

The observed behavior of $A$ and $R$ implies that above the transition and before the dominance of the electroweak sector, the system has essentially the EoS of interacting matter. On the other hand, above $T_c$ the color degrees of freedom are still not neutralized and therefore the previous results suggest the formation of colored and massive clusters near the deconfinement transition before the formation of colorless bound states.

Indeed, this is a well known interpretation of the QCD EoS at finite temperature in terms of quasi-particles where quasi-quarks and quasi-gluons have a dynamical, temperature dependent, effective masses which mimic the interaction and that near $T_c$ are large, i.e. in the range $\simeq 0.6 - 1.2$ GeV for $\mu_B=0=B$ \cite{quasi1,quasi1.0,quasi2,quasi3}.

\section{Comments and Conclusions}

The analysis in the previous sections  shows that during the deconfinement transition the time evolution of the scale factor and of 
$H(t)$ are weakly sensitive to the EoS and that , on the contrary, the cosmological parameters $q(t)$ and $j(t)$ follows the behavior of the ratio $p/\epsilon$. Starting from a radiation dominated Universe, the time evolution of the EoS indicates that above and near the transition time $t_c$ the entire system (quark gluon plasma + electroweak sector) is in a matter dominated state ($q\approx 0.5$ and $j\approx 1$). For longer time, the evolution is again dominated by the radiation EoS. 

The introduction of a finite baryon chemical potential and a background magnetic field do not qualitatively change this dynamical picture.

On the other hand, since above and near $T_c$ one has color degrees of freedom, the matter state which drive the EoS is , presumably, formed by color massive objects, as suggested by the quasi-particle models
\cite{quasi1,quasi1.0,quasi2.0}.
Indeed, the behavior of $A(t)$, i.e. of the matter content of the system, reported in fig. 15 is analogous to the time evolution
of the interaction measure $(\epsilon-3p)/T^4$, in fig. 16, where the electroweak sector has no role.

\begin{figure}
{{\epsfig{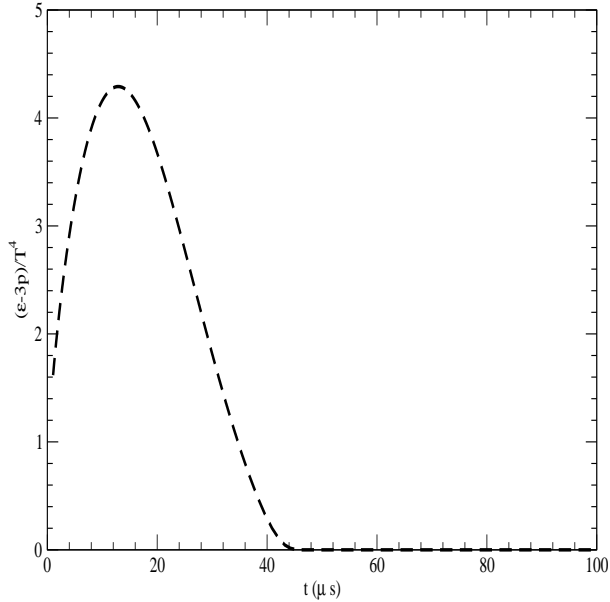}}
\caption{Time evolution of the interaction measure. 
}}
\end{figure}

The effective degrees of freedom below $T_c$, i.e. for $t > t_c$, are not included in the present study and the extrapolations for long time give the time evolution of a radiation dominated Universe. On the other hand, below $T_c$ and at small baryon chemical potential one expects the formation of $q \bar q$ states which decay in electroweak final states. 
It is known that a hadron resonance gas model is able to give a good description of the QCD matter
down to $T\sim 1/3 T_c$ \cite{Ratti:2010kj,Vovchenko:2014pka}. Including properly the EoS of hadronic matter probably will lead to further
extend the stage of matter dominated dynamics even beyond $t\approx \, 100 \rm fm/c$.


\end{document}